\documentclass[prc,showpacs,showkeys,aps,epsfig]{revtex4}

\usepackage{psfig}

\def\lsim{\raise0.3ex\hbox{$<$\kern-0.75em\raise-1.1ex\hbox{$\sim$}}}

\def\gsim{\raise0.3ex\hbox{$>$\kern-0.75em\raise-1.1ex\hbox{$\sim$}}}

\newcommand{\be}{\begin{equation}}

\newcommand{\ee}{\end{equation}}

\def\alphaem{\alpha_{em}}

\def\beq{\begin{equation}}

\def\eeq{\end{equation}}

\def\beqa{\begin{eqnarray}}

\def\eeqa{\end{eqnarray}}

\newcommand{\rr}{\mbox{\boldmath $r$}}

\newcommand{\rb}{\mbox{\boldmath $b$}}

\def\gappeq{\mathrel{\rlap {\raise.5ex\hbox{$>$}}

{\lower.5ex\hbox{$\sim$}}}}

\def\lappeq{\mathrel{\rlap{\raise.5ex\hbox{$<$}}

{\lower.5ex\hbox{$\sim$}}}}

\def\Toprel#1\over#2{\mathrel{\mathop{#2}\limits^{#1}}}

\begin{document}

\title{Heavy quark production in deep inelastic electron-nucleus scattering}
\author{ V.P. Gon\c{c}alves $^{1}$, M.S. Kugeratski $^{2}$ 
and  F.S. Navarra$^3$}
\affiliation{$^{1}$ Instituto de F\'{\i}sica e Matem\'atica,  Universidade
Federal de Pelotas\\
Caixa Postal 354, CEP 96010-090, Pelotas, RS, Brazil\\
$^2$Universidade Federal de Santa Catarina, Campus Universit\'ario de Curitibanos,\\
CEP  89520-000, Curitibanos, SC, Brazil\\
$^3$Instituto de F\'{\i}sica, Universidade de S\~{a}o Paulo,
C.P. 66318,  05315-970 S\~{a}o Paulo, SP, Brazil\\} 

\begin{abstract}

Heavy quark production has been very well studied over the last years both theoretically and 
experimentally. Theory has been used to study heavy quark production in $e p$ collisions at 
HERA, in $pp$ collisions at Tevatron and at RHIC, in $pA$ and $dA$ collisions at RHIC and in 
$AA$ collisions at CERN-SPS and at RHIC. However, to the best of our knowledge, heavy quark 
production in $eA$ has received almost no attention. With the 
possible construction of a high energy electron-ion collider, updated  estimates of heavy 
quark production are needed.  We  address the subject from the perspective of saturation 
physics and  compute the heavy quark production cross section with the dipole model. We  
isolate shadowing and non-linear effects, showing their impact  on the charm structure 
function and on the transverse momentum spectrum.

\end{abstract}



\maketitle

\vspace{1cm}

\section{Introduction}

The construction of a high energy  Electron Ion Collider (EIC) was proposed in 2005 
\cite{erhic} (See also \cite{dainton}). During the subsequent years, several predictions for the inclusive and 
diffractive observables were made, especially in the context of saturation physics. 
One of the advantages of this new machine is that it  will be possible to reach values 
of the saturation scale, $Q_s$,  which are larger than those reached at HERA. A large 
saturation scale is crucial for the observation of most of the saturation effects. In 
particular,  the collider environment is ideal for studying semi-inclusive and exclusive 
processes. In previous works, \cite {kgn1,kgn2,ccgn1,ccgn2,gkmn_mv}, we made predictions
for the inclusive nuclear structure function $F_2$ and $F_L$ as well for the diffractive 
observables. In these works we have concluded that the nuclear structure function $F_2^A$ 
is reduced up to $50\%$ with respect to case where  saturation effects are not taken into 
account. We made estimates for the ratio between the nuclear diffractive and  total cross 
sections and  predicted that about $30\%$ of the events at an EIC will be diffractive. 
We have also investigated the dependence on the  $\beta$ and $ x_{I\!\!P}$ variables of the 
nuclear diffractive structure function $ x_{I\!\!P}F_{2,A}^{D(3)}$. We showed that 
$x_{I\!\!P}F_{2,A}^{D(3)}$ becomes very flat in  $\beta$ and $ x_{I\!\!P}$ when we increase 
the atomic number, $A$, and we found the same behavior for  the ratio 
$R=F_{2,A1}^{D(3)}/F_{2,A2}^{D(3)}$ for two different nuclei. Concerning the exclusive vector 
meson production off nuclei, we showed that the coherent process (when the nucleus remains 
intact after the collision) for vector meson production  will be much more important  than 
the incoherent one (when the nucleus breaks up after the collision).  

In this paper  we continue our study of quantities that could be measured in an 
electron ion collider and calculate the  cross section  of heavy quark production  
using the dipole approach and a nuclear saturation model based on the physics of the  
Color Glass Condensate  (CGC) (For related studies see Refs. \cite{ab,gm}). The main input of  our calculation  is the dipole-nucleus 
cross section, 
$\sigma_{dA} (x,\rr)$,  which is determined by the QCD dynamics at small $x$. In the eikonal 
approximation it is  given by:
\begin{equation} 
\sigma_{dA} (x, \rr) = 2 \int d^2 \rb \,  {\cal N}^A (x, \rr, \rb)
\label{sdip}
\end{equation}
where $ {\cal N}^A (x, \rr, \rb)$ is the forward dipole-target scattering amplitude for a 
dipole with size $\rr$ and impact parameter $\rb$ which encodes all the information about the 
hadronic scattering, and thus about the non-linear and quantum effects in the hadron wave 
function (see e.g. \cite{cgc}). It  can be obtained by solving the BK (JIMWLK) evolution 
equation in the rapidity $Y \equiv \ln (1/x)$ \cite{BAL,KOVCHEGOV,JIMWLK}. Many groups have studied the numerical solutions  
of the BK equation, but several improvements are still necessary  before using the solution in 
the calculation of  observables. In particular, one needs to include the next-to-leading order 
corrections into the evolution equation and perform a global analysis of all 
small $x$ data. It is a program in progress (for recent results see \cite{alba,alba2}). 
In the meantime it is necessary to use phenomenological models for ${\cal N}$ which capture 
the most essential properties of the solution. Following \cite{ccgn2} we will use in our 
calculations  the model proposed in Ref. \cite{armesto}, which describes  the current 
experimental data on the nuclear structure function as well as includes the  impact parameter 
dependence in the dipole nucleus cross section. In this model the forward dipole-nucleus 
amplitude is given by
\begin{eqnarray}
{\cal{N}}^A(x,\rr,\rb) = 1 - \exp \left[-\frac{1}{2}  \, \sigma_{dp}(x,\rr^2) \,T_A(\rb)\right] \,\,,
\label{enenuc}
\end{eqnarray}
where $\sigma_{dp}$ is the dipole-proton cross section and $T_A(\rb)$ is the nuclear profile 
function, which is obtained from a 3-parameter Fermi distribution for the nuclear
density normalized to $A$ (for details see, e.g., Ref. \cite{gm}).
The above equation, based on the Glauber-Gribov formalism \cite{gribov},  sums up all the 
multiple elastic rescattering diagrams of the $q \overline{q}$ pair and is justified for 
large coherence length, where the transverse separation $r$ of partons in the multiparton 
Fock state of the photon becomes a conserved quantity, {\it i.e.} the size of the pair $r$  
becomes eigenvalue of the scattering matrix. It is important to emphasize that for very small 
values of $x$, other diagrams beyond the multiple Pomeron exchange considered here should 
contribute ({\it e.g.} Pomeron loops) and a more general approach for the high density 
(saturation) regime must be considered. However, we believe that the present approach allows 
us to estimate the magnitude of the high density effects in the  kinematical range of the 
future $eA$ colliders. 
\begin{figure}[t]
\vspace{1.0cm}
\centerline{\psfig{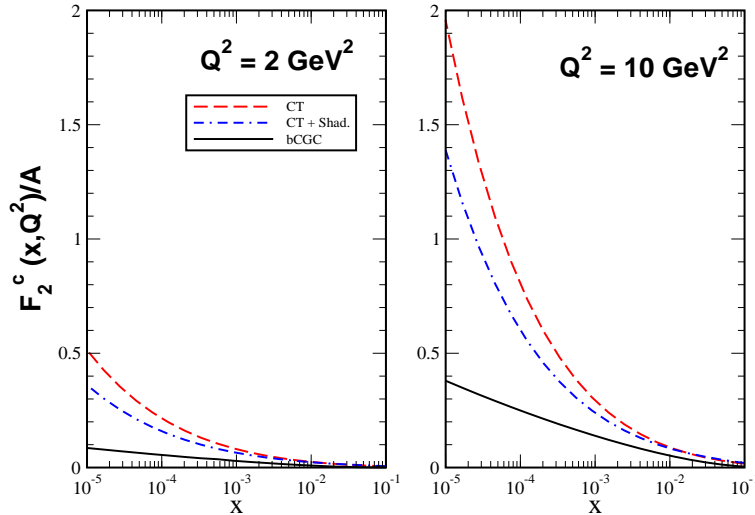}} 
\vspace{0.5cm}
\caption{(Color online) Nuclear charm  structure functions for $A= Pb$ and different values 
of $Q^2$ considering different models for the dipole-nucleus cross section.}
\label{fig1}
\end{figure}

During  the last years an intense activity in the area resulted  in  more 
sophisticated dipole proton cross sections, which had more  theoretical 
constraints and which were able to give a better description of the more recent 
HERA and/or RHIC data \cite{kmw,pesc,watt08,kkt,dhj,gkmn,buw}. In what follows we will use 
the b-CGC model proposed in Ref. \cite{kmw}, which improves the IIM model 
 \cite{iim} with  the inclusion of  the impact parameter dependence in the dipole 
proton cross sections. The parameters of this model were recently fitted to describe 
the current HERA data \cite{watt08}.  Following \cite{kmw} we have that the dipole-proton 
cross section is given by:
\begin{equation}
\sigma_{dp}^{bCGC} (x,\rr^2) \equiv \int \, d^2 \bar{\rb} \, \frac{d 
\sigma_{dp}}{d^2  \bar{\rb}} 
\label{new_iim}
\end{equation}
where 
\begin{eqnarray}
\frac{d \sigma_{dp}}{d^2 \bar{\rb}} = 2\,\mathcal{N}^p(x,\rr,\bar{\rb}) =  2 \times 
\left\{ \begin{array}{ll} 
{\mathcal N}_0\, \left(\frac{ r \, Q_{s}}{2}\right)^{2\left(\gamma_s + 
\frac{\ln (2/r Q_{s})}{\kappa \,\lambda \,Y}\right)}  & \mbox{$r Q_{s} \le 2$} \\
 1 - \exp \big[-a\,\ln^2\,(b \, r \, Q_{s})\big]   & \mbox{$r Q_{s}  > 2$} 
\end{array} \right.
\label{eq:bcgc}
\end{eqnarray}
with  $Y=\ln(1/x)$ and $\kappa = \chi''(\gamma_s)/\chi'(\gamma_s)$, where $\chi$ is the 
LO BFKL characteristic function.  The coefficients $a$ and $b$  
are determined uniquely from the condition that $\mathcal{N}^p(x,\rr)$  and its 
derivative 
with respect to $rQ_s$  are continuous at $rQ_s=2$. They are given by:
\begin{eqnarray}
a = - \frac{{\cal{N}}_0^2 \gamma_s^2}{(1-{\cal{N}}_0)^2\ln(1-{\cal{N}}_0)} \,\,\,\,\mbox{and} \,\,\,\, b = \frac{1}{2}(1-{\cal{N}}_0)^{-\frac{(1-{\cal{N}}_0)}{{\cal{N}}_0 \gamma_s}}\,\,.
\end{eqnarray}
 
In this model, the proton saturation scale $Q_{s}$ now depends on the impact 
parameter:
\begin{equation} 
  Q_{s}\equiv Q_{s}(x,\bar{\rb})=\left(\frac{x_0}{x}\right)^{\frac{\lambda}{2}}\;
\left[\exp\left(-\frac{\bar{b}^2}{2B_{\rm CGC}}\right)\right]^{\frac{1}{2\gamma_s}}.
\label{newqs}
\end{equation}
The parameter $B_{\rm CGC}$  was  adjusted to give a good 
description of the $t$-dependence of exclusive $J/\psi$ photoproduction.  
Moreover the factors $\mathcal{N}_0$ and  $\gamma_s$  were  taken  to be free. In this 
way a very good description of  $F_2$ data was obtained. 
The parameter set  which is going to be used here is the one presented in the second 
line of Table II of \cite{watt08}:  $\gamma_s = 0.46$, $B_{CGC} = 7.5$ GeV$^{-2}$,
$\mathcal{N}_0 = 0.558$, $x_0 = 1.84 \times 10^{-6}$ and $\lambda = 0.119$.

In order to estimate the magnitude of the saturation effects in  heavy quark production 
it is important to compare  the CGC predictions  with those associated to linear QCD 
dynamics. As a model for the linear regime we consider the leading logarithmic approximation 
for the dipole-target cross section \cite{nik_dip,fran_raz}, where $\sigma_{dA}$ is directly 
related to the nuclear gluon distribution $xg_A$ as follows
\begin{eqnarray}
\sigma_{dA}(x,\rr^2) =  {\frac{\pi^2}{3}} \rr^2 \alpha_s xg_A(x, 10/\rr^2) \,\,.
\label{gluongrv}
\end{eqnarray}
The use of this cross section in the formulas given below will produce results which we
denote CT, from color transparency. In this limit we are disregarding multiple scatterings of the dipole with the nuclei and are assuming that the dipole interacts incoherently with the target. In what follows we consider two different models for 
the nuclear gluon distribution. In the first one we disregard the nuclear effects and 
assume that $xg_A (x,Q^2) = A.xg_N (x,Q^2)$, with $xg_N$ being the  gluon distribution in 
the proton and given by the GRV98 parameterization \cite{grv98}. We will refer to this 
model as CT. In the second model we take into account the nuclear effects in the nuclear 
gluon distribution as described by the EKS98 parameterization \cite{eks}. In this case 
$xg_A (x,Q^2) = A.R_g(x,Q^2).xg_N (x,Q^2)$ with $R_g$ given in \cite{eks}. We will call this 
model CT + Shad. In our calculations the charm quark mass is $m_c = 1.5$ GeV and the  bottom 
quark  mass is $m_b = 4.5$ GeV.

\section{Heavy Quark Production}
\label{sec:nc}

The electron-proton ($ep$) collider at HERA has opened up a new kinematic regime in the study of the deep structure of the proton and, in general, of hadronic interactions, which is characterized by small values of the Bjorken variable $x= Q^2 / s$, where $Q^2$ is the momentum transfer and $\sqrt{s}$ is the center-of-mass energy. In this regime we expect that the usual collinear approach \cite{collfact} be replaced by a more general factorization scheme, as for example the   $k_{\perp}$-factorization  approach \cite{CCH,CE,GLRSS} or the quasi-multi-Regge-kinematics (QMRK) framework \cite{lip_action} (For related studies see Refs. \cite{ic,ic_kniehl,ic_kop}). Let us present a brief review of these distinct approaches.

In the collinear factorization approach \cite{collfact} all partons involved are assumed to be on mass shell, carrying only longitudinal momenta, and their transverse momenta are neglected in the QCD matrix elements. Moreover, the cross sections for the QCD subprocess are usually calculated in the leading order (LO), as well as in the next-to-leading order (NLO). In particular, the cross sections involving incoming hadrons are given, at all orders, by the convolution of intrinsically non-perturbative (but universal) quantities - the parton densities - with perturbatively calculable hard matrix elements, which are process dependent. The conventional gluon distribution $g(x,\mu^2)$, which drives the behavior of the observables at high energies, corresponds to the density of gluons in the proton having a longitudinal momentum fraction $x$ at the factorization scale $\mu$. This distribution satisfies the DGLAP evolution in $\mu^2$ and does not contain information about the transverse momenta $k_{\perp}$ of the gluon. On the other hand, in the large energy (small-$x$) limit, we have that the characteristic scale $\mu$ of the hard subprocess of parton scattering is much less than $\sqrt{s}$, but greater than the $\Lambda_{QCD}$ parameter. In this limit, the effects of the finite transverse momenta of the incoming partons become important, and the factorization must be generalized, implying that the cross sections are now $k_{\perp}$-factorized into an off-shell partonic cross section and a  $k_{\perp}$-unintegrated parton density function ${\cal{F}}(x,k_{\perp})$, characterizing the $k_{\perp}$-factorization  approach \cite{CCH,CE,GLRSS}.  The function $\cal{F}$ is obtained as a solution  of the   evolution equation associated to the dynamics that governs the QCD at high energies.  
A sizeable piece  of the NLO and some of the NNLO corrections to the LO contributions on the collinear approach, related to  the  contribution of non-zero transverse momenta of the incident partons, are already included in the LO contribution within the $k_{\perp}$-factorization approach. Moreover, the coefficient functions and the splitting functions giving the collinear parton distributions are supplemented by all-order $\alpha_s\ln (1/x)$ resummation at high energies \cite{CH}. A detailed  comparison between the predictions of the collinear and $k_{\perp}$-factorization approaches for the heavy-quark photoproduction was performed in Refs. \cite{gm,Mariotto_Machado}, which we indicate for more details of these two approaches. 

In the last years, an alternative approach to calculated the heavy quark production at high energies was proposed considering the QMRK  framework. It  is based on an effective theory implemented with the non-Abelian gauge-invariant action obtained in Ref. \cite{lip_action}. In this approach the initial-state $t$-partons are considered as  Reggeons. In contrast to the $k_{\perp}$-factorization approach, the QMRK approach uses gauge-invariant amplitudes and is based on a factorization hypothesis that is proven in the leading logarithmic approximation. 
The phenomenological implications of this approach were discussed in detail in Refs. \cite{saleev1,saleev2,saleev_hq}, which demonstrated that the QMRK approach is a powerful tool for the theoretical description of the high energy processes. In particular, in \cite{saleev_hq} the $F_2^c$ and $D$-meson spectra are successfully described using the QMRK approach.

The heavy quark production  can also be calculated using the  color dipole approach \cite{nik_dip}. This formalism can be obtained from the $k_{\perp}$-factorization approach  after the Fourier transformation from the space of quark transverse	 momenta into the space of transverse coordinates (See e.g. \cite{predazzi}). It is important to emphasize that this equivalence is only valid in the leading logarithmic approximation, being violated if the exact gluon kinematics is considered \cite{navelet}. A detailed discussion of the equivalence or not  between the dipole  and the QMRK approaches  still is an open question (See, however, Refs. \cite{fiore}).  The main advantage to use the color dipole formalism, is that it gives a simple unified picture of inclusive and diffractive processes and the saturation effects can be easily implemented in this approach. It is important to emphasize that 
phenomenological models based on the Color Glass Condensate (See, e.g., \cite{watt08}) or the solution of the running coupling BK equation \cite{rcbk,vjm,vma}  describe quite well the current experimental  HERA data for inclusive and exclusive observables.

\subsection{Charm structure function}

In terms of  virtual photon-target cross sections  $\sigma_{T,L}$
for the transversely and longitudinally  polarized photons, the nuclear  
$F_2$ structure function is given by
$$ 
F_2^A(x,Q^2)\,=\,\frac{Q^2}{4 \pi^2 \alphaem} \,\sigma_{tot}(x,Q^2)
$$ 
with \cite{nik_dip}:
\be
\label{eq:1}
\sigma_{tot} = \sigma_T\,+\,\sigma_L \,\,\,\mbox{and}\,\,\,\sigma_{T,L}\,=\,  
\int d^2{\rr}\, dz\, |\Psi_{T,L}(\rr,z,Q^2)|^2\,\, \sigma_{dA} (x,\rr),
\ee
where $\Psi_{T,L}$ is the light-cone  wave function of the virtual photon
and $\sigma_{dA}$ is the   dipole nucleus cross section
describing  the interaction of the $q\bar{q}$  dipole with the nucleus target.  In
Eq. (\ref{eq:1})
$\rr$ is the transverse separation of the $q\bar{q}$ pair
and $z$ is the photon  momentum fraction carried by the quark (for details see  
e.g. Ref. \cite{predazzi}). The charm component of the 
nuclear structure function $F_2^{c,A}(x,Q^2)$ is obtained directly from 
Eq. (\ref{eq:1}) isolating the charm flavor. In Fig. \ref{fig1} we show $F_2^{c}(x,Q^2)/A$. 
As expected, in this kinematical domain it grows with $Q^2$ and falls with increasing $x$.
What is really  remarkable is the difference  between the bCGC and the linear CT models, 
which can reach a factor up to 4!.  In previous estimates of this observable \cite{cdns},  
non-linear effects were found to be  weaker. However, in \cite{cdns}  the input was different
(unintegrated gluon distribution instead of a dipole cross section) and the procedure 
adopted to estimate the purely linear contribution was to switch off the non-linear effects in 
the unintegrated gluon distribution of the proton. In some of our previous works 
(for example in \cite{gkmn_mv})  we adopted an analogous procedure and tried to make this 
separation switching off the non-linear component of the dipole-proton  cross section. 
Although this method could give us some rough idea of the role played by some non-linear 
effects, we were missing part of  them associated with the fusion of gluons belonging to 
different nucleons. Therefore we believe that the mentioned previous estimates have 
underestimated the importance of non-linear effects.

\subsection{Heavy quark spectrum}

As discussed before, heavy quark production  has been very well studied over the last years both theoretically and 
experimentally. The elementary cross sections have been calculated in perturbative QCD up 
to next lo leading order and the parton densities have been extracted with the same degree 
of precision. Theory has been used to study heavy quark production in $e p$ collisions at 
HERA, in $pp$ collisions at Tevatron and at RHIC, in $pA$ and $dA$ collisions at RHIC and in 
$AA$ collisions at CERN-SPS and at RHIC. A recent and comprehensive survey of these advances 
can be found in \cite{fuv}. However, to the best of our knowledge, heavy quark production in 
$eA$ has received almost no attention. This is probably related to the fact that $eA$ data 
are old and until recently, there was no prospect of having high energy $eA$ data. With the 
possible construction of a high energy EIC, updated  estimates of heavy quark production are 
needed.  We wish to address the subject from the perspective of saturation physics and thus 
the best option to obtain the production cross section, isolating shadowing and non-linear 
effects, is to use  the dipole model. The  dipole approach  is very natural for the study of 
exclusive hidden charm and beauty  electro and photo-production especially in the vector 
meson channel. As it was shown in  \cite{floter} the dipole formalism can be easily extended to 
open charm and beauty electro-production obtaining a quite successful description of the HERA data for the $F_2^c$ and $D$-meson spectra.  Here, in order to calculate the differential heavy 
quark production cross section, $d\sigma^{T,L}/d^{2}p_{Q}^{\bot}$, we have extended the 
approach of  Ref. \cite{floter}, which was originally developed for $ep$ scattering,  
to electron-ion collisions with the Glauber-Gribov formalism. 
In this extension we implicitly assume that the factorization of the cross section verified for $ep$ collisions remains valid in the nuclear case and make use of the dipole-nucleus cross section, which, in turn, 
depends on the dipole-nucleon cross section. For this last quantity we take the recent 
parametrization given by Eqs. (\ref{new_iim}) and  (\ref{eq:bcgc}). The resulting cross section 
reads:
\begin{eqnarray}  
\frac{d\sigma(\gamma^{*}A\rightarrow Q\,X)}{d^{2}p_{Q}^{\bot}}  
&=&\frac{6e_{Q}^{2}\alpha_{em}}{(2\pi)^{2}}\int d\alpha
 \left\lbrace\left[\vphantom{\frac{1}{1}}m_{Q}^{2}+
4Q^{2}\alpha^{2}(1-\alpha)^{2}\right]  
\right.\left[\frac{I_{1}}{p_{Q}^{\bot 2}+\epsilon^{2}}-\frac{I_{2}}{4\epsilon}\right]
\nonumber\\  
&+&\left[\vphantom{\frac{1}{1}}\alpha^{2}+(1-\alpha)^{2}\right]\left.  
\left[\frac{p_{Q}^{\bot}\epsilon I_{3}}{p_{Q}^{\bot 2}+  
\epsilon^{2}}-\frac{I_{1}}{2}+\frac{\epsilon I_{2}}{4}\right]\right\rbrace  
\label{eqn:epI}  
\end{eqnarray}  
with   
\begin{eqnarray}  
I_{1}&=&\int dr\,r\,J_{0}(p_{Q}^{\bot}r)\,K_{0}(\epsilon r)\,\sigma_{dA}(\rr)\nonumber\\  
I_{2}&=&\int dr\, r^{2}\, J_{0}(p_{Q}^{\bot}r)\,K_{1}(\epsilon r)\,\sigma_{dA}(\rr)
\nonumber\\  
I_{3}&=&\int dr\, r\, J_{1}(p_{Q}^{\bot}r)\,K_{1}(\epsilon r)\,\sigma_{dA}(\rr)\ .  
\label{eqn:I1-I3}  
\end{eqnarray}  
where $J_{0,1}$ and $K_{0,1}$ are Bessel functions, 
$\epsilon = \alpha (1 - \alpha) Q^2 + m^2$ and $\sigma_{dA}$ is given by Eq. (\ref{sdip}) or (\ref{gluongrv}).

To calculate the  $D$-meson production cross section we must let the charm quark fragment. 
Following  \cite{floter} we convolute the charm quark production cross section 
(\ref{eqn:epI}) with the nonperturbative fragmentation function:
\begin{eqnarray}
&&\frac{d\sigma(\gamma^{*}A\rightarrow DX)}{dz\,d^{2}p_{D}^{\bot}}  
=\int\frac{dp_{c}^{\bot}\,d\alpha}{\alpha}\,  
\frac{d\sigma(\gamma^{*}A\rightarrow cX)}{d^{2}p_{c}^{\bot}\,d\alpha} \, D^{c}_{D}\left(\frac{z}{\alpha}\right) \delta\left(p^{\bot}_{D}-\frac{z}  
{\alpha} p^{\bot}_{c}\right)\, , 
\label{eqn:conv. mit z}
\end{eqnarray}
where $D_{Q}^{h}(z^{*})$ is the well known Peterson fragmentation function given by
\begin{eqnarray}
D_{Q}^{h}(z^{*})&=&\frac{n(h)}{z^{*}[1-\frac{1}{z^{*}}-
\frac{\epsilon_{Q}}{1-z^{*}}]^{2}}\, \,.
\label{eqn:pe} 
\end{eqnarray}
The fragmentation function gives the probability that the original charm quark with a 
momentum $P$ fragments into a $D$-meson with momentum fraction $z^{*} P$. There are more
recent fragmentation functions but here we are only  interested in checking  if the 
differences between linear and non-linear dynamics are affected by fragmentation. For this 
purpose the Peterson fragmentation function is adequate.

\begin{figure}[t]
\centerline{{\psfig{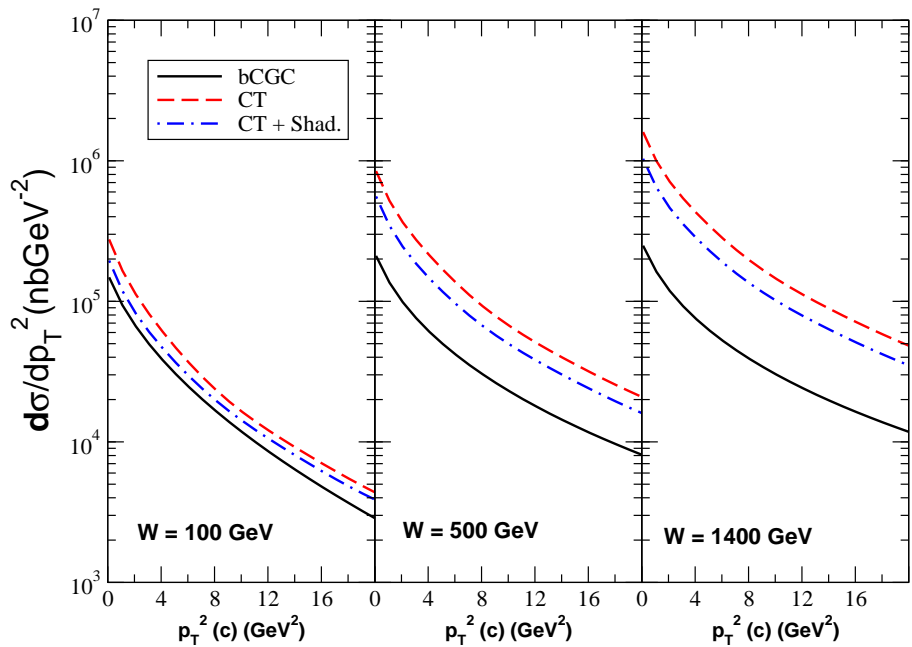}}}
\caption{(Color online) Transverse momentum charm spectrum for $Q^2 = 2$ GeV$^2$ and different 
center-of-mass energies.}
\label{fig2}
\end{figure}

In Fig. \ref{fig2} we show the transverse momentum spectrum of charm quarks. The main 
purpose of this figure is to show that the predictions of the  linear physics (CT + Shad) 
differ from the total  (i.e. bCGC)  by a factor which increases with 
the energy $W$ and goes from  $1.5$ ($W=100$ GeV) to $4$ ($W=1400$ GeV). Moreover, this 
difference persists for a wide momentum window. At very large $p_{T}$ we enter the deep 
linear  regime and expect that the two curves coincide. 

\begin{figure}[t]
\vspace{0.5cm}
\centerline{{\psfig{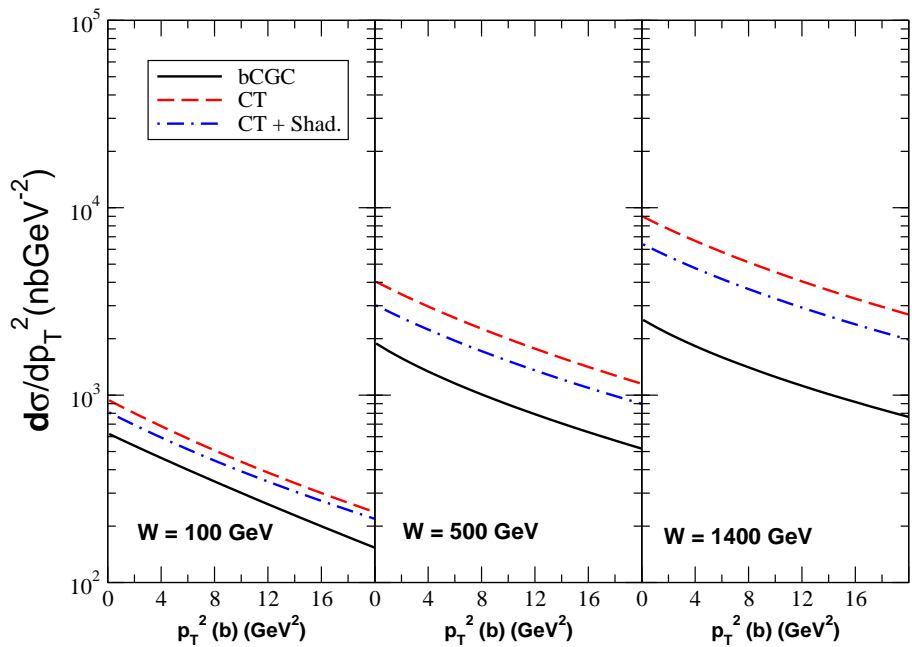}}}
\caption{(Color online) Transverse momentum bottom spectrum for $Q^2 = 2$ GeV$^2$ and different 
center-of-mass energies.}
\label{fig3}
\end{figure}

\begin{figure}[t]
\vspace{0.5cm}
\centerline{{\psfig{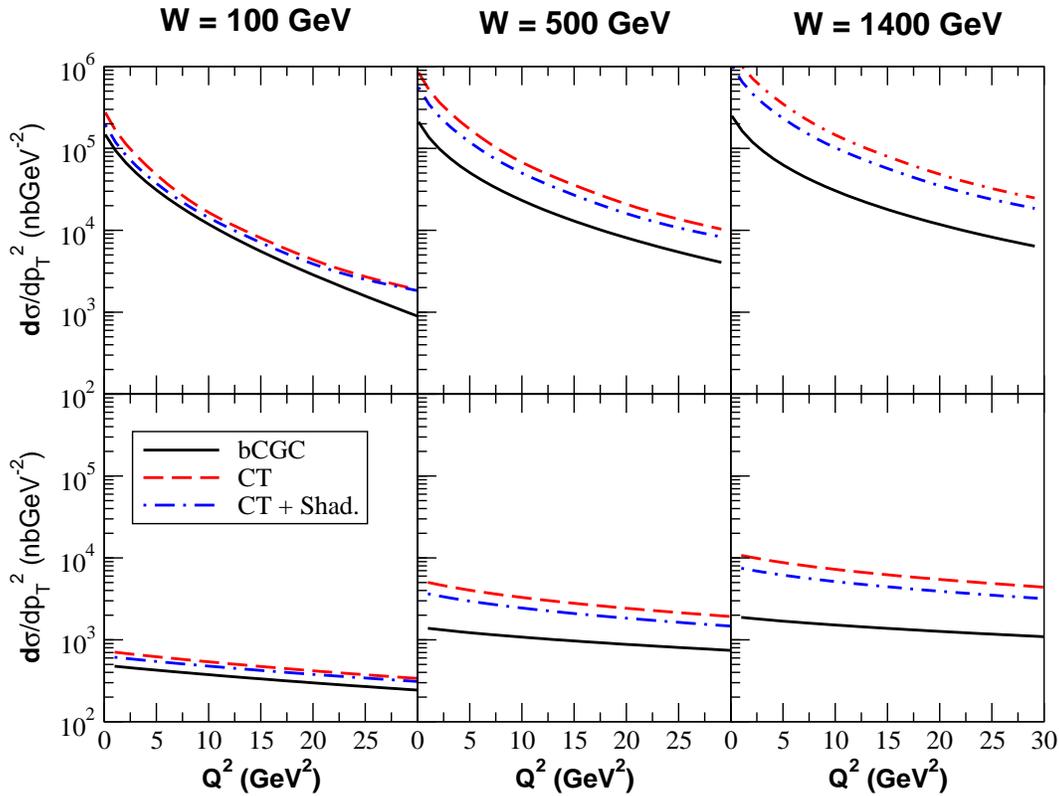}}}
\vspace{0.5cm}
\caption{(Color online) Dependence on the photon virtuality at $p_T^2 = 4$ GeV$^2$.}
\label{fig4}
\end{figure}

\begin{figure}[t]
\vspace{0.5cm}
\centerline{{\psfig{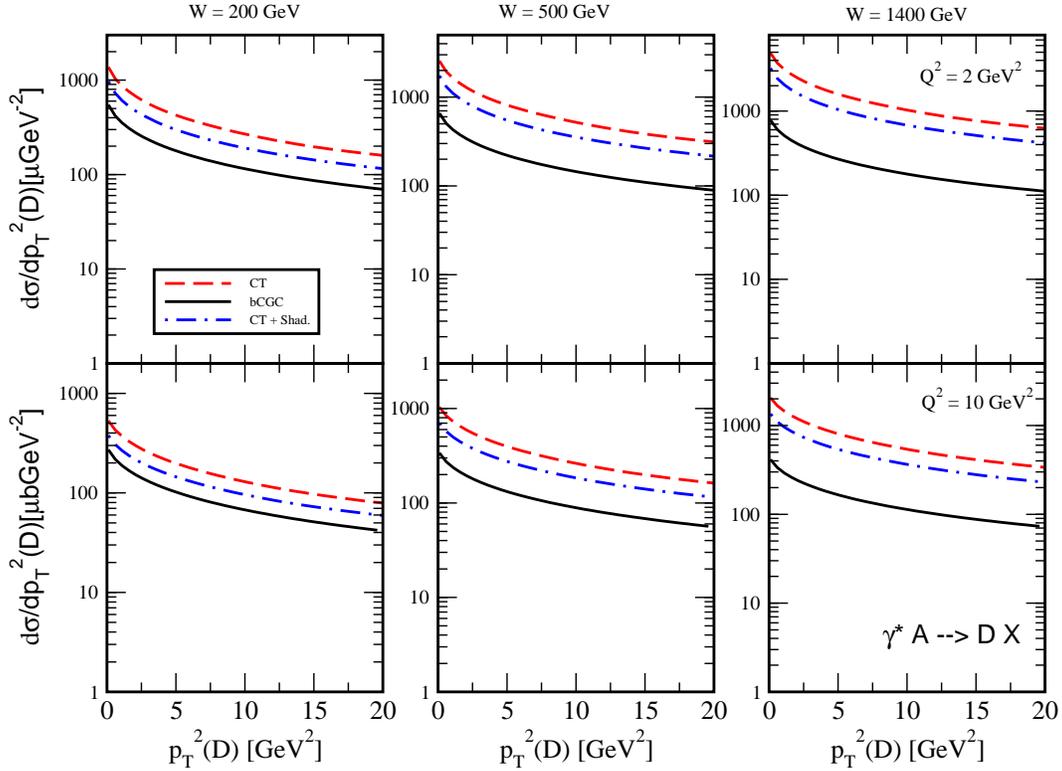}}}
\vspace{0.5cm}
\caption{(Color online) Transverse momentum spectra of $D$ mesons.}
\label{fig5}
\end{figure}

In Fig. \ref{fig3} we show the transverse momentum spectrum of bottom quarks. As expected, 
we observe the same features of the charm distribution, except that now the non-linear 
effects are weaker. Nevertheless they are still noticeable. 
In Fig. \ref{fig4} we show the $Q^2$ dependence of the $p_{T}$ distribution at a fixed 
value $p_{T} = 4$ GeV$^2$ for different energies. The upper and lower panels show the charm 
and bottom distributions respectively. Here again, we observe a remarkable strenght and 
persistence up to large virtualities of the differences between CT + Shad and bCGC.  
In Fig. \ref{fig5}   we show the transverse momentum spectrum of $D$ mesons for 
three energies $W= 200 , 500 , 1400$ GeV  and for two virtualities $Q^2 = 2$ GeV$^2$ (upper 
panels) and  $Q^2 = 10$ GeV$^2$ (lower panels). As it can be seen,  the differences between 
the curves CT, CT +Shad and bCGC are the same as before. 

\section{Conclusions}
\label{sec:conc}

In this work we have  updated the calculations presented in 
\cite{floter} and extended them to electron-ion collisions. We compared the 
predictions of a saturation model (bCGC) with the predictions made with a
linear model. The main conclusion was that it seems quite possible to observe 
the non-linear effects both in the charm structure function  $F_2^{c,A}(x,Q^2)$ 
and  in the $p_{T}$ distributions of the heavy quarks.  For the energies considered 
this difference is of a factor going from $1.5$ to $4$. As expected, the final state 
conversion of the heavy quarks into heavy mesons, performed through the convolution of 
our $p_{T}$ distributions with the Peterson fragmentation function, does not change the 
difference between ``full'' (= linear + non-linear) and linear predictions. In a future 
analysis of fragmentation we shall include the production of heavy mesons from light 
quarks. Although the  fragmentation channel $q \rightarrow H$ (where $q$ is a light quark 
and $H$ is a D or B meson) is disfavoured in comparison with $Q \rightarrow H$, the production 
of light quarks from the incoming photon is strongly enhanced by the photon wave function. 
One effect might compensate the other and, in the end, light quarks might play a significant 
role in heavy meson production \cite{kopef}. If this would be the case, non-linear effects 
would be even stronger. 

Our results suggest that heavy quark production in high energy $eA$ collisions is a 
promising signature of saturation. Previous estimates of this same observable were 
not so positive, probably because they addressed $ep$ or $pp$ collisions as in 
Ref. \cite{peters} or because the method employed to separate linear and non-linear  
effects was not very accurate. From our figures we can also conclude that non-linear 
dynamics, here as in several other contexts, leads to a depletion in the $p_T$ spectra, 
in contrast to some early estimates \cite{dainese}.

A final comment is in order. As discussed in Section II, the heavy quark production at high energies can be calculated considering different approaches which are not  equivalent in the full kinematic region. Consequently, a more detailed study of the saturarion effects using these different approaches is important in order to estimate the theoretical uncertainty of our predictions. It is postponed for a future publication. 
\begin{acknowledgments}

This work was  partially financed by the Brazilian funding 
agencies CNPq and FAPESP.

\end{acknowledgments}

\hspace{1.0cm}


\begin{thebibliography}{99}


\bibitem{erhic} A. Deshpande, R. Milner, R.
               Venugopalan and  Volgelsang, { Ann. Rev. Nucl. Part. Sci.} {\bf 55}, 165 (2005).

\bibitem{dainton}
  J.~B.~Dainton, M.~Klein, P.~Newman, E.~Perez and F.~Willeke,
  JINST {\bf 1}, P10001 (2006)


\bibitem{kgn1}  M.~S.~Kugeratski, V.~P.~Goncalves and F.~S.~Navarra, 
               Eur.\ Phys.\ J.\  C {\bf 46}, 413 (2006).

  
\bibitem{kgn2}  M.~S.~Kugeratski, V.~P.~Goncalves and F.~S.~Navarra,  
               Eur.\ Phys.\ J.\  C {\bf 46}, 465 (2006).


\bibitem{ccgn1}   E.~R.~Cazaroto, F.~Carvalho, V.~P.~Goncalves and F.~S.~Navarra, 
                  Phys.\ Lett.\  B {\bf 669}, 331 (2008). 


\bibitem{ccgn2}   E.~R.~Cazaroto, F.~Carvalho, V.~P.~Goncalves and F.~S.~Navarra, 
                  Phys.\ Lett.\  B {\bf 671}, 233 (2009). 


\bibitem{gkmn_mv}  V.~P.~Goncalves, M.~S.~Kugeratski, M.~V.~T.~Machado and F.~S.~Navarra,  
                   Phys.\ Rev.\ C {\bf 80}, 025202 (2009). 

\bibitem{ab}
  N.~Armesto and M.~A.~Braun,
  Eur.\ Phys.\ J.\  C {\bf 22}, 351 (2001)



\bibitem{gm}
  V.~P.~Goncalves and M.~V.~T.~Machado,
  Eur.\ Phys.\ J.\  C {\bf 30}, 387 (2003)




\bibitem{cgc} E.~Iancu and R.~Venugopalan,  arXiv:hep-ph/0303204;
              A.~M.~Stasto,  Acta Phys.\ Polon.\ B {\bf 35}, 3069 (2004);
              H.~Weigert,  Prog.\ Part.\ Nucl.\ Phys.\  {\bf 55}, 461 (2005);
              J.~Jalilian-Marian and Y.~V.~Kovchegov,
              Prog.\ Part.\ Nucl.\ Phys.\  {\bf 56}, 104 (2006); F.~Gelis, E.~Iancu, J.~Jalilian-Marian and R.~Venugopalan,
  arXiv:1002.0333 [hep-ph].


\bibitem{BAL}  I. I. Balitsky,   Nucl. Phys. {\bf  B463}, 99 (1996); Phys. Rev. Lett. {\bf 81}, 2024 (1998); Phys. Rev. D  {\bf 60}, 014020 (1999).


\bibitem{KOVCHEGOV}  
Y.V. Kovchegov,  Phys. Rev. D {\bf 60},  034008 (1999);  Phys. Rev. D {\bf 61} 074018 (2000). 

\bibitem{JIMWLK} J. Jalilian-Marian, A. Kovner, L. McLerran  and  H. 
Weigert, Phys. Rev. D {\bf 55}, 5414  (1997); 
J. Jalilian-Marian, A. Kovner and  H. 
Weigert, Phys. Rev. D {\bf 59}, 014014  (1999); Phys. Rev. D{\bf 59}, 014015  (1999); Phys. Rev. D {\bf 59}, 034007 (1999); A. Kovner, J. Guilherme Milhano and  H. Weigert,   Phys. Rev. D {\bf 62}, 114005 
(2000).
 


\bibitem{alba}
  J.~L.~Albacete,
  Phys.\ Rev.\ Lett.\  {\bf 99}, 262301 (2007). 

  

\bibitem{alba2} J.~L.~Albacete, N.~Armesto, J.~G.~Milhano and C.~A.~Salgado, 
                Phys.\ Rev.\  D {\bf 80}, 034031 (2009).
  

 
\bibitem{armesto} N.~Armesto,
  Eur.\ Phys.\ J.\  C {\bf 26}, 35 (2002).






 






  

\bibitem{gribov}
V. N. Gribov, Sov. Phys. JETP {\bf 29}, 483 (1969); Sov. Phys. JETP {\bf 30}, 709 (1970).

 
 

\bibitem{kmw}
  H.~Kowalski, L.~Motyka and G.~Watt,
  Phys.\ Rev.\  D {\bf 74},  074016 (2006). 



\bibitem{pesc}
  C.~Marquet, R.~B.~Peschanski and G.~Soyez,
  Phys.\ Rev.\  D {\bf 76}, 034011 (2007)



\bibitem{watt08} 
G.~Watt and H.~Kowalski,
  Phys.\ Rev.\  D {\bf 78}, 014016 (2008).





\bibitem{kkt} D. Kharzeev, Y.V. Kovchegov and K. Tuchin,
              { Phys. Lett.} {\bf  B599}, 23 (2004).


\bibitem{dhj}
  A.~Dumitru, A.~Hayashigaki and J.~Jalilian-Marian,
  Nucl. Phys. {\bf A765}, 464 (2006); Nucl.Phys. {\bf A770}, 57 (2006).


\bibitem{gkmn}
  V.~P.~Goncalves, M.~S.~Kugeratski, M.~V.~T.~Machado and F.~S.~Navarra,
  Phys.\ Lett.\  B {\bf 643}, 273 (2006). 


\bibitem{buw}   D.~Boer, A.~Utermann and E.~Wessels,
  Phys.\ Rev.\  D {\bf 77},  054014 (2008).



\bibitem{iim} E. Iancu, K. Itakura, S. Munier,
                   Phys. Lett. B {\bf 590}, 199  (2004).



\bibitem{nik_dip}  N.~N.~Nikolaev and B.~G.~Zakharov, Z. Phys. {\bf  C49}, 
 607 (1991); Z. Phys. {\bf C53}, 331  (1992);  A.~H.~Mueller, Nucl. Phys.
{\bf B415}, 373  (1994); A.~H.~Mueller and B.~Patel, Nucl. Phys. {\bf B425}, 471 
(1994).  

\bibitem{fran_raz}  L.~Frankfurt, A.~Radyushkin and M.~Strikman, 
                    Phys.\ Rev.\  D {\bf 55}, 98 (1997).


\bibitem{grv98}   M.~Gluck, E.~Reya and A.~Vogt, 
                  Eur.\ Phys.\ J.\  C {\bf 5}, 461 (1998).


\bibitem{eks} K.~J.~Eskola, V.~J.~Kolhinen and P.~V.~Ruuskanen, 
              Nucl.\ Phys.\  B {\bf 535}, 351 (1998);
              K.~J.~Eskola, V.~J.~Kolhinen and C.~A.~Salgado,
              Eur.\ Phys.\ J.\  C {\bf 9}, 61 (1999). 


\bibitem{collfact}
J.C. Collins, D.E. Soper, G. Sterman,   Factorization of hard processes in QCD.  In: Mueller, A. H. (Ed.) $\,\,$ {\it 
Perturbative quantum chromodynamics.} $\,\,$ Singapore: World Scientific, 
1989.


\bibitem{CCH}
S.~Catani, M.~Ciafaloni, F.~Hautmann,  Nucl. Phys. {\bf B366}, 135 (1991).

\bibitem{CE}
J.~Collins, R.~Ellis, Nucl. Phys. {\bf B360}, 3 (1991).

\bibitem{GLRSS}
L.~Gribov, E.~Levin, M.~Ryskin,  Phys. Rep. {\bf 100}, 1 (1983); \\
E.M. Levin, M.G. Ryskin, Y.M. Shabelski, A.G. Shuvaev, Sov. J. Nucl.
  Phys. {\bf 53}, 657 (1991).


\bibitem{lip_action}
  L.~N.~Lipatov,
  Nucl.\ Phys.\  B {\bf 452}, 369 (1995); V.~S.~Fadin and L.~N.~Lipatov,
  Nucl.\ Phys.\  B {\bf 477}, 767 (1996);  E.~N.~Antonov, L.~N.~Lipatov, E.~A.~Kuraev and I.~O.~Cherednikov,
  Nucl.\ Phys.\  B {\bf 721}, 111 (2005).
  

\bibitem{ic}
V. P. Goncalves, F. S. Navarra, T. Ullrich, Nucl.\ Phys.\  A {\bf 842}, 59 (2010).

\bibitem{ic_kniehl}
B.~A.~Kniehl, G.~Kramer, I.~Schienbein and H.~Spiesberger,
  Phys.\ Rev.\  D {\bf 79}, 094009 (2009)

\bibitem{ic_kop}
B.~Z.~Kopeliovich, I.~K.~Potashnikova and I.~Schmidt,
arXiv:1003.3673 [hep-ph].
\bibitem{CH}
S.~Catani and F.~Hautmann, Nucl. Phys. B {\bf 427}, 475 (1994).



\bibitem{Mariotto_Machado}
C. Brenner Mariotto, M.B. Gay Ducati, M.V.T. Machado,  Phys. Rev. D {\bf 66}, 114013 (2002).


\bibitem{saleev1}
 B.~A.~Kniehl, D.~V.~Vasin and V.~A.~Saleev,
  Phys.\ Rev.\  D {\bf 73}, 074022 (2006); Phys.\ Rev.\  D {\bf 74}, 014024 (2006)



\bibitem{saleev2}
 V.~A.~Saleev,
  Phys.\ Rev.\  D {\bf 78}, 034033 (2008); Phys.\ Rev.\  D {\bf 78}, 114031 (2008)



\bibitem{saleev_hq}
B.~A.~Kniehl, A.~V.~Shipilova and V.~A.~Saleev,
  Phys.\ Rev.\  D {\bf 79}, 034007 (2009)

  







\bibitem{predazzi}  V.~Barone and E.~Predazzi,
\textit{High-Energy Particle Diffraction}, Springer-Verlag, Berlin Heidelberg, (2002).


\bibitem{navelet}
A.~Bialas, H.~Navelet and R.~B.~Peschanski,
  Nucl.\ Phys.\  B {\bf 603}, 218 (2001)



\bibitem{fiore}
V.~S.~Fadin, R.~Fiore and A.~Papa,
  Nucl.\ Phys.\  B {\bf 769}, 108 (2007); Phys.\ Lett.\  B {\bf 647}, 179 (2007); Nucl.\ Phys.\  B {\bf 784}, 49 (2007)

\bibitem{rcbk}
  J.~L.~Albacete, N.~Armesto, J.~G.~Milhano and C.~A.~Salgado,
  Phys.\ Rev.\  D {\bf 80}, 034031 (2009)


\bibitem{vjm}
  M.~A.~Betemps, V.~P.~Goncalves and J.~T.~de Santana Amaral,
  Eur.\ Phys.\ J.\  C {\bf 66}, 137 (2010)

\bibitem{vma}
  V.~P.~Goncalves, M.~V.~T.~Machado and A.~R.~Meneses,
   Eur.\ Phys.\ J.\  C ({\it in press});  arXiv:1003.0828 [hep-ph].



\bibitem{cdns}  F.~Carvalho, F.~O.~Dur\~aes, F.~S.~Navarra and S.~Szpigel,
  Phys.\ Rev.\  C {\bf 79}, 035211 (2009).



\bibitem{fuv} A.D. Frawley, T. Ullrich, R. Vogt, Phys. \ Rept.\  {\bf 462},  125 (2008). 


\bibitem{floter} B.~Fl\"oter,  B. Z. Kopeliovich, H.-J.~Pirner and J.~Raufeisen, 
                         Phys. Rev. D {\bf 76}, 014009 (2007).

\bibitem{kopef} We thank B. Z. Kopeliovich for instructive discussions on this subject. 


\bibitem{peters} K. Peters, hep-ph/0606265. 

\bibitem{dainese}  A. Dainese, R. Vogt, M. Bondila, K.J. Eskola, V.J. Kolhinen, 
                   J. Phys. G  {\bf 30}, 1787 (2004). 



\end{thebibliography}
\end{document}